# Enhanced betatron X-rays from axially modulated plasma wakefields

J.P. Palastro, D. Kaganovich, and D. Gordon
*Naval Research Laboratory, Washington DC 20375-5346, USA*

## Abstract

In the cavitation regime of plasma-based accelerators, a population of high-energy electrons trailing the driver can undergo betatron motion. The motion results in X-ray emission, but the brilliance and photon energy are limited by the electrons' initial transverse coordinate. To overcome this, we exploit parametrically unstable betatron motion in a cavitated, axially modulated plasma. Theory and simulations are presented showing that the unstable oscillations increase both the total X-ray energy and average photon energy.

When a relativistic charged particle beam or intense laser pulse driver displaces background electrons from its path, it leaves behind a region of net positive charge with dimensions approximating the plasma wavelength [1-10]. A witness electron beam following the driver will be attracted to the center of this region, and undergo transverse and longitudinal oscillations or betatron motion [11-17]. As a result of the acceleration, the witness electrons emit Larmor radiation in the form of X-rays. Any experimental system capable of wakefield acceleration is capable of producing betatron radiation, making it a convenient and promising X-ray source.

Two shortcomings of betatron radiation from wakefield accelerators are the lack of power radiated into X-rays and the achievable X-ray frequencies. As discussed by Ta Phuoc *et al.* [13], there are several approaches to overcoming this: increasing the witness beam energy, increasing the background plasma density, or increasing the transverse excursion of the electron. Using simulations, Ta Phuoc *et al.* showed that a single divot or ramp placed in the background electron density increases the maximum transverse excursion, and, as a result, the power radiated at higher frequencies.

Here we consider electrons undergoing betatron oscillations in modulated plasmas [18-21]. When the modulation period is matched to one half the betatron period, a parametric resonance occurs where the electron's betatron amplitude grows exponentially. Consequently, the radiated power increases *exponentially* with time, the radiation spectrum broadens, and the average X-ray frequency grows proportionally.

A relativistic charged particle beam or intense laser pulse driver travelling through plasma displaces plasma electrons. In the 'blow-out' or 'bubble' regime, the driver completely expels the plasma electrons from its path leaving behind a bare ion cavity with dimensions approximating the plasma wavelength, $\lambda_p \sim 2\pi/k_p$, where $k_p = 4\pi e^2 n_i / m_e c$, $n_i$ is the number density of positive charges, $e$ is the elementary charge, $m_e$ the electron mass, and $c$ the speed of light [22-25]. Modeling the system as an ion cavity, or electron density hole, moving along the z-axis near the speed of light, Kostyuokuv *et al.* found the electrostatic and vector potentials inside the cavity to be

$$\Phi = -\frac{1}{8}(\xi^2 + x^2 + y^2) \quad (1a)$$

$$A_z = -\Phi, \quad (1b)$$

where $\xi = z - v_0 t$ measures distance in a frame moving at $v_0$ with the drive beam and cavity and the coordinate system is defined with the center of the cavity at $(\xi, x, y) = (0, 0, 0)$ [22]. In Eq. (1) and the remainder of the manuscript, potentials, distances, and times are normalized by $e/m_e c^2$, $k_p$, and $\omega_p = k_p/c$ respectively.

Here we consider a plasma with axial density modulations [18-21] in the bubble regime. The charge number densities upstream from the drive beam are modeled as $n_e = n_i = n_0[1 - \delta \cos(k_m z)]$, where $k_m = 2\pi/\lambda_m$, $\lambda_m$ is the modulation length, and $0 \leq \delta \leq 1$. The cavity formed behind the drive beam is devoid of electrons such that $n_e = 0$ and $n_i = n_0[1 - \delta \cos(k_m z)]$. If the modulation length is much longer than the plasma wavelength, $\lambda_m \gg 1$, the electrostatic and vector potentials are approximately

$$\Phi \simeq -\frac{1}{8}(x^2 + y^2 + \xi^2)[1 - \delta \cos(k_m z)] + \frac{1}{16}\delta k_m \xi \left(x^2 + y^2 + \frac{2}{3}\xi^2\right)\sin(k_m z) \quad (2a)$$

$$A_z = -\Phi. \quad (2b)$$

In the limit $\beta_0 = v_0/c \to 1$, Eq. (2) results in $\nabla \cdot \mathbf{E} = 4\pi[1 - \delta \cos(k_m z)]$ and zero current density, $\mathbf{J} = 0$, to order $k_m^2$. In the following we will drop corrections to the coefficients in Eq. (2) of order $\gamma_0^{-2} = 1 - \beta_0^2$.

A relativistic witness electron beam trailing the drive beam will evolve in response to the potentials. In particular, the potentials provide transverse and longitudinal restoring forces which will initiate oscillations for off-center witness beam electrons. The equations of motion for the witness beam electrons are

$$\frac{d\mathbf{x}}{dt} = \frac{\mathbf{P}}{\gamma} \quad (3a)$$

$$\frac{dP_x}{dt} = (1 + v_z)\frac{\partial}{\partial x}\Phi \quad (3b)$$

$$\frac{dP_z}{dt} \simeq -\left(\frac{\partial}{\partial t} - \frac{\partial}{\partial z} + v_x \frac{\partial}{\partial x}\right)\Phi, \quad (3c)$$

where $\gamma = [1 + \mathbf{P} \cdot \mathbf{P}]^{1/2}$ and only the transverse motion in the $\hat{\mathbf{x}}$-direction is considered.

For an approximate solution to Eqs. (3), we perform a series expansion of the dynamic variables in inverse orders of $\gamma_0 = (1 - \beta_0^2)^{-1/2}$, where $\beta_0 = v_0/c$ and $v_0$ is the initial axial velocity of the witness beam electrons. For an electron starting at the center of the cavity, we have to second order $x(t) = x_1(t)$ and $\xi(t) = z_2(t)$ which evolve according to

$$\frac{d^2 x_1}{dt^2} = -\omega_\beta^2 [1 - \delta \cos(\omega_m t)] x_1 \quad (4a)$$

$$\frac{d^2 z_2}{dt^2} \simeq \frac{1}{2} \omega_\beta^2 [1 - \delta \cos(\omega_m t)] \frac{d}{dt} x_1^2, \quad (4b)$$

where $\omega_m = k_m$ and $\omega_\beta = k_\beta = (2\gamma_0)^{-1/2}$ is the betatron frequency. In the absence of modulations, $\delta = 0$, the solution to Eq. (4) is the well known betatron trajectory of electrons in a bare ion cavity [7,8]:

$$x_1(t) = x_\beta \cos(\omega_\beta t) \quad (5a)$$

$$z_2(t) \simeq -\frac{1}{8} k_\beta x_b^2 \left[ 2\omega_\beta t - \sin(2\omega_\beta t) \right], \quad (5b)$$

where $x_\beta$ is the initial displacement of the electron. The electrons undergo transverse oscillations with constant amplitude equal to their initial distance from the axis.

With modulations, Eq. (4) supports parametric resonances when the curvature of the potential well oscillates in phase with the electron, ie. $k_m = 2k_\beta$, $k_m = k_\beta$, etc. In standard betatron motion the electron undergoes harmonic motion with the restoring force increasing as the electron approaches its minimum or maximum transverse excursion. When $k_m = 2k_\beta$, the increase in restoring force is reduced, allowing the electron to reach a larger transverse excursion before stopping and reversing direction at the end of each half cycle. Such a parametric instability, in which the vector potential of a laser pulse provides the modulation, was recently exploited by Arefiev et al. for increasing electron energy [26].

Using a multi-scale technique to solve Eq. (4) with $k_m = 2k_\beta$, one can show that the electron trajectory is reasonably approximated by

$$x_1(t) \simeq \frac{1}{2^{1/2}} x_\beta \left[ e^{\frac{1}{4}\delta\omega_\beta t} + (2^{1/2} - 1) \right] \cos(\omega_\beta t) \quad (6a)$$

$$z_2(t) \simeq -\frac{1}{4} k_\beta x_\beta^2 \left( \frac{e^{\frac{1}{2}\delta\omega_\beta t} - 1}{\delta} + \frac{1}{2}\omega_\beta t \right) + \frac{1}{16} k_\beta x_\beta^2 (1 + e^{\frac{1}{2}\delta\omega_\beta t}) \sin(2\omega_\beta t) \quad (6b)$$

for up to 10 betatron periods when $0 \leq \delta < 0.3$. Equations (6a) shows that the transverse excursion of the electron grows exponentially with a rate $\delta\omega_\beta / 4$ while its axial velocity decreases exponentially: $\beta_{z2} \simeq -(k_\beta x_\beta / 2)^2 (1 + e^{\frac{1}{2}\delta\omega_\beta t}) \sin^2(\omega_\beta t)$. Figure 1 compares our analytical solution to the numerical solution of Eqs. (2) and (3) for $k_\beta = 0.022$, $x_\beta = 0.1$, $k_m = 0.045$, and $\delta = 0.25$. The analytical solution is in good agreement with the numerical solution for several hundred plasma periods. The exponentiation of the betatron oscillation amplitude is readily observed, increasing by a factor of ~20. Additionally, the electron slips backwards in the speed of light frame consistent with the exponential decrease in axial velocity.

The total power radiated by a single electron can be found from the Larmor formula [27]

$$P(t) = \frac{2}{3} r_e \gamma^6 \left[ (\dot{\boldsymbol{\beta}})^2 - (\boldsymbol{\beta} \times \dot{\boldsymbol{\beta}})^2 \right], \quad (7)$$

where $r_e$ is the classical electron radius. Using Eq. (6a) with Eq. (7), the cycle averaged power is

$$P(t) \simeq P_* x_\beta^2 [1 + (1-\delta)^2] \left[ e^{\frac{1}{4}\delta\omega_\beta t} + (2^{1/2} - 1) \right]^2, \quad (8)$$

where $P_* = r_e \gamma_0^2 / 48$. The temporal exponentiation of the electron's acceleration translates to an exponential enhancement in the radiated power and energy when compared to uniform plasmas. Figure 2 compares the power radiated during betatron motion in uniform, top, and modulated plasmas, bottom. The dashed lines are Eq. (8), while the solid lines are Eq. (7) evaluated with numerical solutions to Eqs. (2) and (3) for the same parameters above. The radiated power peaks every half betatron period when the electron changes directions at its largest transverse excursion. Integrating the radiated power, we find that the betatron oscillations in the modulated plasma result in ~14 times more radiated energy than the uniform plasma over ~300 plasma periods. Over longer times, the enhancement could be even larger so long as non-ideal effects do not spoil the motion. We will consider such effects later.

Of more interest for applications is the forward radiated power and spectrum. The radiated energy spectrum per unit solid angle, $\Omega$, can be calculated from

$$\frac{d^2U}{d\omega d\Omega} = \frac{r_e \omega^2}{4\pi^2} \left| \int dt \left[ \mathbf{n} \times (\mathbf{n} \times \boldsymbol{\beta}) \right] \exp\left[ i\omega(t - \mathbf{n} \cdot \mathbf{x}/c) \right] \right|^2 \quad (9)$$

where $\mathbf{n}$ is the unit vector pointing to the location of observation [27]. As above, we consider transverse motion in the $\hat{\mathbf{x}}$ direction only. Integrating Eq. (9) over the azimuth, $\phi$, we find the forward radiated energy spectrum:

$$\left. \frac{d^2U}{d\omega d\Theta} \right|_{\theta=0} = \frac{r_e \omega^2}{2\pi} \left| \int_0^T dt \beta_x \exp\left[ i\omega(t - z/c) \right] \right|^2, \quad (10)$$

where $d\Theta = \sin\theta d\theta$. Figure 3 displays the forward radiated spectrum calculated from the numerical solution to Eqs. (2) and (3) as a function of frequency for $\delta$=0.0, 0.25, and 0.5. All other parameters are the same as above. The horizontal axis is in units of the resonant radiation frequency $\omega_r = 2^{1/2} \gamma_0^{3/2} [1 + \gamma_0 x_\beta^2 / 4]^{-1}$ [11], and the spectrum has been Gaussian filtered with a width of $\omega_r$. The $\delta$=0.0 and 0.25 spectra were calculated over an interval $T = 1800$, while the $\delta$=0.5 interval was limited to $T = 900$. The $\delta$=0.5 interval was limited so that the final betatron oscillation amplitude was equal to that of $\delta$=0.25 at $T = 1800$. Such a situation represents the $\delta$-dependent time required for transverse ejection of electrons from the bubble. We return to this below.

When $\delta$=0.0 the spectrum exhibits the standard $\omega_r$ X-ray harmonics [11]. For non-zero delta the spectrum becomes broadened with the higher frequencies becoming enhanced. The broadening results from three effects. First, the temporal exponentiation of the betatron amplitude provides a resonance broadening for each $\omega_r$-harmonic. For short times ($T \ll 2/\delta\omega_\beta$), we can replace $x_\beta^2$ by $(1 + \delta\omega_\beta T / 8)x_\beta^2$ in the expression for $\omega_r$. This results in a red-shifting of the radiated harmonics and reduces their spacing causing an accumulation of the harmonics at lower frequencies. Numerical calculations not presented in this manuscript confirm this effect is stronger for large $\delta$ and times. Finally, the contribution of higher betatron harmonics become more pronounced for larger $\delta$. This results in a proliferation of the resonances allowed by Eq. (10): $\omega(t-z/c) - n\omega_\beta \approx 0$. An analysis of the transition from a spectrum consisting of well-defined resonances in a uniform plasma to a thermal-like spectrum in a modulated plasma requires a level of detail beyond the scope of the current manuscript.

Surprisingly, Fig. 3 shows that a $\delta=0.25$ results in more forward X-ray emission than $\delta=0.5$. Recall that the intervals for calculating the spectra were chosen such that the maximum betatron amplitude was approximately equal for both cases. Furthermore, calculations similar to those presented in Fig. 2 show slightly greater total power radiated for $\delta=0.5$, $T=900$. The apparent reduction in the forward emitted X-rays has two causes. First, calculations (not presented) show that the spectrum for $\delta=0.5$ contains more energy in the frequency range extending beyond $100\omega_r$ than the $\delta=0.25$ spectrum. Second, the fraction of energy radiated to larger angles increases with $\delta$. The on-axis radiation is driven predominately by the motion along x, while the lateral radiation is driven by motion along z. From Eq. (5b), we see that $\beta_{z2} \propto \exp[\delta\omega_\beta t/2]$ increases disproportionately with $\delta$ relative to $\beta_{x1} \propto \exp[\delta\omega_\beta t/4]$, increasing the relative contribution of the off-axis radiation to the total radiated energy. Optimizing the betatron radiation, angle, and spectrum by tuning $\delta$ and other laser and plasma parameters will be a focus of future work.

After integrating over frequencies in the range 0 to 200 $\omega_r$, the on-axis energy is ~3.0 times larger in the $\delta=0.25$ plasma than in the uniform plasma. We note this is a conservative estimate: the spectrum in the modulated case decays slowly with increasing frequency. For an example of an experimental realization, Eq. (9) must be integrated over the acceptance angle of the X-ray detector. Employing a small angle approximation in the forward direction we have

$$\left.\frac{dU}{d\omega}\right|_{det} = \frac{r_e\omega^2}{4\pi}\theta^2\left|\int_0^T dt\beta_x \exp[i\omega(t-z/c)]\right|^2. \quad (11)$$

To estimate the total number of photons, we divide the total energy, $U = \int (dU/d\omega)|_{det} \, d\omega$, by the average photon energy, $<\hbar\omega> = U^{-1} \int \hbar\omega (dU/d\omega)|_{det} \, d\omega$. For $n_0 = 1 \times 10^{18}$ cm$^{-3}$ and $\gamma_0 = 1000$ with the remaining parameters the same as above, we find $N_p = 3900 N_e \theta^2$ and $<\hbar\omega> = 12$ keV for $\delta = 0.0$ and $N_p = 5300 N_e \theta^2$ and $<\hbar\omega> = 25$ keV for $\delta = 0.25$, where $N_e$ is the number of electrons undergoing betatron motion.

Several effects can limit the continued exponentiation of the betatron amplitude. As alluded to in regards to Fig. 3, the maximum transverse excursion will be limited by the size of the ion cavity, $r_b$, and thus the number of betatron oscillations will be limited to $N_\beta \sim (2/\pi)\delta^{-1} \ln(2^{1/2} r_b / x_\beta)$. As the electron undergoes increasingly larger transverse oscillations, its axial velocity decreases exponentially: $\beta_{z2} \simeq -(k_\beta x_\beta / 2)^2 (1 + e^{\frac{1}{2}\delta\omega_\beta t}) \sin^2(\omega_\beta t)$ causing it to dephase with respect to the bubble. When the transverse excursion becomes large, the electron will experience spatial anharmonicities in the bubble potential which may spoil the motion. Finally, any evolution of the driver could cause the bubble potential to evolve in time.

To demonstrate the unstable betatron motion persists in the presence of these effects, we perform ponderomotive guiding center (PGC) simulations of laser-driven cavitation with TurboWAVE [28,29]. The simulations are conducted in cylindrical coordinates preserving the transverse dynamics of the cavitation. The laser pulse was initialized with a sin$^2$ temporal profile, and a Gaussian transverse profile. The pulse parameters were as follows: central wavelength $\lambda = 0.15$, temporal full width half maximum (FWHM) $\tau = 2$, exp(-1) field width $w = 4$, and normalized amplitude, $a_0 = eA_\perp / m_e c^2 = 4$, where $A_\perp$ is the transverse vector potential of the laser pulse. The plasma wavenumber is the same as above: $k_m = 0.045$, while $\delta = 0.9$. In principle electrons self-injected into the laser-driven plasma wave can undergo unstable betatron oscillations. For a proof of concept, we load a beam of test particles into the bubble behind the pulse. The test particle beam has negligible charge, a length $L = 2.5$, width

$W = 0.3$, $k_\beta = 0.022$, $x_\beta = 0.63$. We use a larger $\delta$ and $x_\beta$ than above to both illustrate the dynamics more clearly and to examine the non-perturbative regime of the instability.

Figure 4 displays the results of the simulations. The normalized charge density, square of the laser amplitude, and test particle density weighted by transverse position (for clarity in display) are plotted in the r-(z-ct) plane at three different axial locations spanning one betatron period. The top and bottom rows illustrate the evolution in axially modulated and uniform plasmas respectively. At z = 1027 the betatron amplitude in the modulated plasma is ~6 times larger than the uniform plasma and limited by the bubble sheath. For later times, not shown, the witness electrons are transversely ejected from the bubble. Similar laser pulse evolution occurs in both plasmas, but the bubble structure is distinct. Specifically, in the modulated plasma, the bubble length undergoes periodic expansions and contractions injecting large amounts of background charge. This can be observed in the bright, on-axis modifications to the charge density. The space-charge fields of the injected charge have scattered the back of the witness beam, spoiling the betatron motion. However, the middle and front of the witness beam exhibit the enhanced transverse excursions indicative of the parametric instability. The betatron period in both the modulated and uniform plasmas varies along the witness beam. This is to be expected as the transverse restoring force depends on the longitudinal coordinate. The parametric instability appears to be relatively insensitive to this source dephasing.

For an experimental realization, one would need to create a pre-formed plasma. Modulated plasmas can be formed by using a ring grating or spatial light modulator (SLM) to impart radial modulations onto a ~100ps Nd:YAG pulse and then axicon focusing the pulse onto a gas jet [18,30]. The axicon maps the radially intensity modulations to axial intensity modulations. The modulated on-axis intensity, ionizes the gas, and heats the plasma, driving hydrodynamic expansion. The resulting plasma profile reflects the non-uniform heating, having high densities where the heater pulse intensity was low and low densities where it was high. Alternatively, one could place a periodic wire array over the gas jet, and by-pass the ring grating or SLM [18,30]. Using these techniques a pre-formed plasma with an average density $n_0$=1x10$^{18}$ cm$^{-3}$, modulation period of 740 µm, and δ=0.25 can be created. Using a laser pulse of 800 nm wavelength, a FWHM of 35 fs, and spot size $w$ = 70 µm, and a 500 MeV electron beam, one could recreate the $N_p = 5300 N_e \theta^2$ scaling with $<\hbar\omega> = 25$ keV found above without coming close to the bubble radius: the maximum betatron amplitude would be ~10 μm, while the bubble radius is ~20 μm.

We have introduced a method of enhancing both the frequency and number of photons emitted during betatron motion in plasma-based accelerators. By modulating the plasma density, a parametric instability was exploited that exponentiated the betatron amplitude. We developed the vector and scalar potentials for ion cavities (bubbles) driven in modulated plasmas. With the potentials, analytic solutions were developed for the betatron motion and the cycle-averaged radiated power. These solutions were validated with numerical solutions to the equations of motion. Forward radiated spectra, showed that unstable betatron motion results in larger total radiated energy and larger photon energies than standard betatron motion. Self-consistent, PGC simulations demonstrated that the betatron instability persists in the presence of driver evolution, background plasma evolution, and anharmonicities in the potentials. We stress that because the enhanced motion results from an instability, it does not require a finely tuned modulation period: the range of detunings, $k_m - 2k_\beta$, for which the instability occurs, increases with δ. Further investigation is required for optimization.

## Acknowledgements

The authors would like to thank A. Arefiev, H.M. Milchberg, and T.M. Antonsen for fruitful discussions. J.P. Palastro would also like to thank the University of Maryland, College Park where a portion of this work was started. This work was supported by the Naval Research Laboratory 6.1 Base Program.

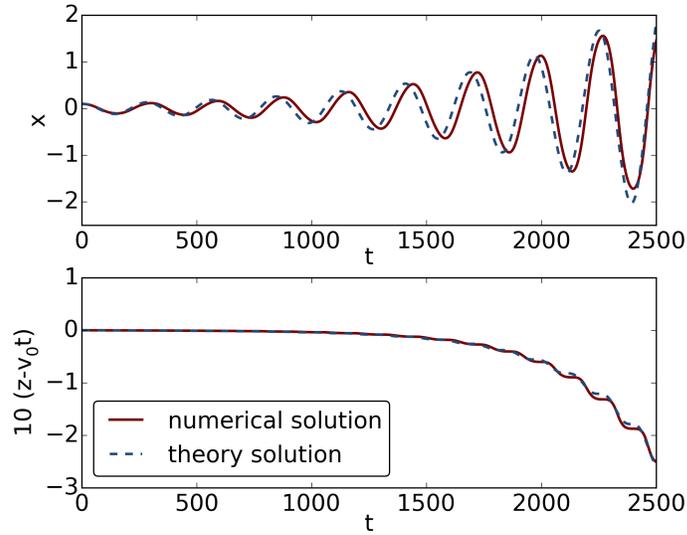

Figure 1 Comparison of the analytic solution in Eq. (6) with the numerical solution to Eq. (3) with the potentials in Eq. (2) for $k_\beta = 0.022$, $x_\beta = 0.1$, $k_m = 0.045$, and $\delta = 0.25$. The temporal exponentiation of the transverse coordinate is readily observed.

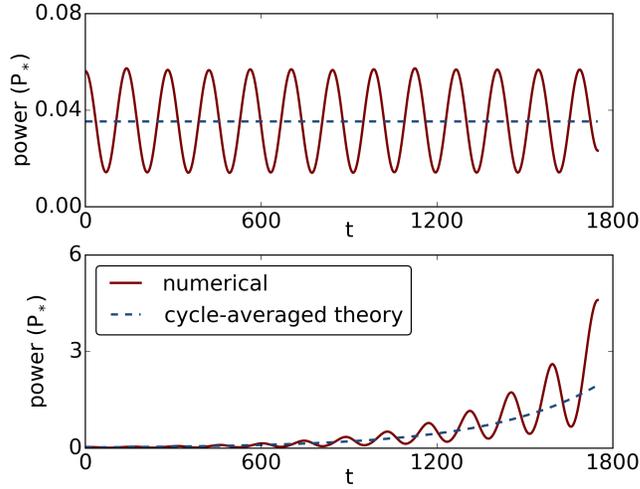

Figure 2 Comparison of the radiated power in a uniform, top, and modulated, bottom, plasma for $k_\beta = 0.022$, $x_\beta = 0.1$, $k_m = 0.045$, and $\delta = 0.25$. The dashed line is Eq. (8) calculated from our analytic solution and the solid line Eq. (7) calculated with the numerical solutions to Eq. (3) with the potentials of Eq. (2). As with the transverse coordinate in Fig. 1, the temporal exponentiation of the radiated power in the modulated plasma is readily observed.

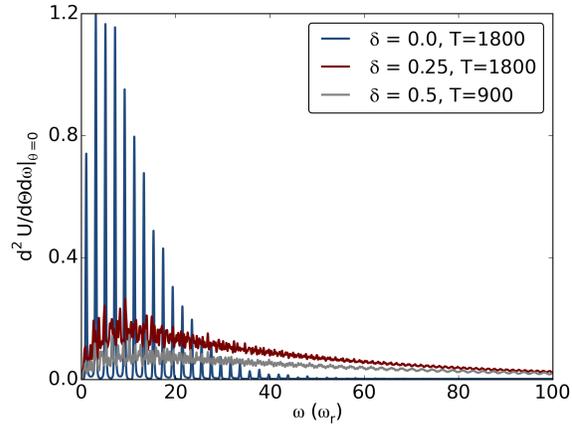

Figure 3 The forward emitted X-ray energy spectra for $k_\beta = 0.022$, $x_\beta = 0.1$, $k_m = 0.045$, and three different modulation amplitudes: δ=0.0 (blue), 0.25 (red), and 0.5 (grey). The time intervals for δ=0.25 and 0.5 were chosen such that their maximum betatron amplitudes were approximately equal.

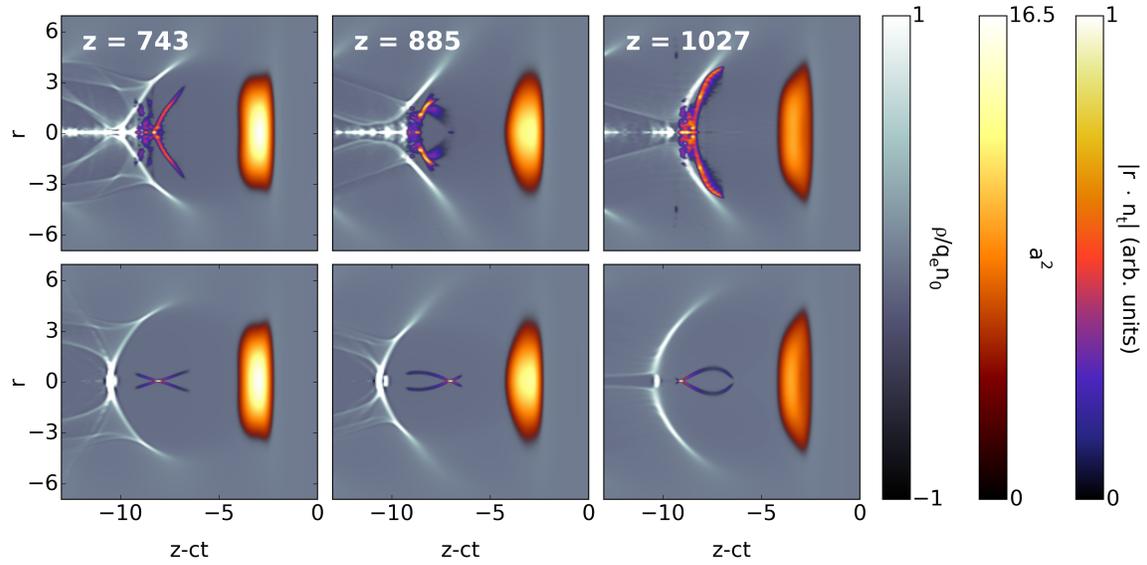

Figure 4 Normalized charge density, square of the laser amplitude, and test particle density weighted by transverse position (for clarity in display) at three different axial distances in the r-(z-ct) plane. The top and bottom rows are for axially modulated and uniform plasmas respectively. The three frames from left to right span one betatron period. At z = 1027 the betatron amplitude in the modulated plasma is ~6 times larger than the uniform plasma and limited by the bubble sheath.